\documentclass{elsart}
\usepackage{epsfig}
\usepackage{amsmath}

\usepackage{amssymb}

\newcommand{\myint}[1]{\int #1 \;d\boldsymbol{x}}
\newcommand{\norm}[1]{\left\Vert #1\right\Vert}

\newtheorem{theorem}{Theorem}

\begin{document}

\begin{frontmatter}

\title{Collapse in coupled Nonlinear Schr\"odinger equations: 
Sufficient conditions and Applications.}
\author{Vadym Vekslerchik}
\ead{vadym.vekslerchik@uclm.es}
\author{Vladyslav Prytula},
\ead{vladyslav.prytula@uclm.es}
\author{V\'{\i}ctor M. P\'erez-Garc\'{\i}a},
\ead{victor.perezgarcia@uclm.es}

\address{Departamento de Matem\'aticas, and Instituto de Matem\'atica Aplicada a la Ciencia y la Ingenier\'{\i}a, E. T. S. de Ingenieros Industriales, 
Universidad de Castilla-La Mancha 13071 Ciudad Real, Spain}


\begin{abstract}
In this paper we study blow-up phenomena in general coupled nonlinear Schr\"odinger equations with different dispersion coefficients. We find sufficient conditions for blow-up and for the existence of global solutions. We discuss several applications of our results to heteronuclear multispecies Bose-Einstein condensates and to degenerate boson-fermion mixtures.
\end{abstract}

\end{frontmatter}

\section{Introduction}

The study of collapse in physical systems governed by scalar nonlinear Schr\"o\-din\-ger (NLS) equations has been a fruitful field in the last 40 years \cite{Berge,Sulem}. Between the physical systems for which the NLS equation is relevant to describe concentration phenomena leading to the formation of a singularity in the mathematical model we can cite the optical self-focusing \cite{self-focusing,Kelley}, Langmuir waves in plasmas \cite{Sulem}, and more recently matter waves, specifically Bose-Einstein condensates \cite{Hulet,Hulet2,F3}.

In other physical situations the field modelled by a Schr\"odinger equation is coupled to another field 
or the physical field itself is of vector nature. Examples of physical systems of this type in which collapse occurs are the Schr\"odinger-Poisson systems \cite{Castella}, the Zakharov equations \cite{Sulem}, the generalized Zakharov model \cite{Pry}, the vector Helmholtz equation \cite{Fibich1,Fibich2,Fibich3}, the Schr\"odinger-Debye system \cite{Besse1} and the Davey-Stewartson system \cite{Besse2}.

However, probably the simplest and most ubiquously arising extension of the scalar nonlinear Sch\"odinger equation is the vector NLS, given by the following equations
\begin{equation}
  i\partial_{t} \psi_{k} = 
  - \frac{1}{2}\lambda_k \Delta \psi_{k} + V_k \psi_k +  \left( \sum_{l} g_{kl} \left|\psi_{l}\right|^{2} \right) \psi_k,
\label{nlse-multi}
\end{equation}
where $\psi_k$, $k=1,...,M$ is a family of $M$ unknown complex functions defined on $\mathbb{R}^D$, $D$ being the spatial dimensionality, $\lambda_k$ are the positive constant dispersion coefficients, $V_k$ are external potentials acting on the different components and $g_{kl}$ are real numbers. 

Collapse in the framework of vector NLS equations has been discussed mainly  for systems of two coupled equations (corresponding to $M=2$)
 and $\lambda_1 = \lambda_2$ in different physical contexts such as plasma beat-wave accelerators
 \cite{McKinstrie} and incoherently coupled waves in bulk Kerr optical media \cite{Bang}. The more general situation of an arbitrary number of components 
has been considered in Ref. \cite{Berge2} for the case $\lambda_k = \lambda$
 in the context of copropagating nonlinear optical waves. The same result was rediscovered by Ghosh \cite{Ghosh} using symmetry methods under similar assumptions (i.e. equal dispersions) in his study of the same model with applications to the dynamics of spinor Bose-Einstein condensates. A more abstract analysis of the same problem with only two components has been recently published \cite{Pomponio} and, in the case of the absence of potential in \cite{Fanelli}.

However, several new scenarios arise in the context of collapses of matter waves, which do not fit into the classes of previously studied  problems. The first one is heteronuclear multicomponent Bose-Einstein condensates \cite{Modugno} where the atomic species have different masses contrary to the case of spinor BECs where different spin states of the same atomic gas are studied. In the language of Eq. (\ref{nlse-multi}) this means that the ``dispersion" coefficients for $\psi_1$ and $\psi_2$ are different, i.e. $\lambda_1 \neq \lambda_2$ \cite{Symbiotic}.

The second field of applications of Eq. (\ref{nlse-multi}) in the theory of degenerate quantum gases is when a mixture of a Bose-Einstein condensate with a degenerate Fermi gas is studied in the mean field limit. These mixtures have received a lot of theoretical attention in the last years, in particular in relation with the description of the observed collapse phenomena \cite{col1,col2,col3}.  One of the models describing the dynamics of boson-fermion mixtures  (see e.g.  Refs. \cite{karpiuk,salerno,Maruyama,solibf}) including the dynamics of the fermionic component is
 \begin{subequations}
 \label{NLSx}
\begin{eqnarray}
i \frac{\partial \psi_B(\boldsymbol{x},t)}{\partial t} & = &  \left[-\frac{1}{2m_B} \Delta  + V_B 
+  g_{BB} |\psi_B(\boldsymbol{x},t)|^2\right] \psi_B(\boldsymbol{x},t)  \nonumber \\ & & 
 + g_{BF} \left(\sum_{j=1}^{M_F} |\phi_j(\boldsymbol{x},t)|^2\right)\psi_B(\boldsymbol{x},t), \label{NLS1}
\\
\label{NLS2}
i  \frac{\partial \phi_j(\boldsymbol{x},t)}{\partial t}  &= &  \left(-\frac{1}{2m_F} \Delta+
V_F+ g_{BF}  |\psi_B(\boldsymbol{x},t)|^2 \right)\phi_j(\boldsymbol{x},t),
\end{eqnarray}
\end{subequations}
where $\phi_j(\boldsymbol{x},t), j=1,...,M_F$ are wavefunctions describing each of the $M_F$ fermions and $\psi(\boldsymbol{x},t)$ is the mean field wavefunction for the bosonic component. 
These equations are written in adimensional units, which are related to the physical parameters as 
$g_{BB} = 4\pi\hbar^2/m_B$, $g_{BF} = 2\pi\hbar^2 (1/m_B+1/m_F)$, $V_F$ and $V_B$ are the magnetic trapping potentials for fermions and bosons respectively, $m_B$ and $m_F$ are 
the masses of the ultracold bosonic and fermionic atoms respectively. The normalizations are 
$\int_{\mathbb{R}^n} |\phi_j(\boldsymbol{x},t)|^2d\boldsymbol{x} = 1,  \int_{\mathbb{R}^n} |\psi(\boldsymbol{x},t)|^2d\boldsymbol{x} = N_B$. The obtention of rigorous criteria for blow-up in this systems is specially interesting since the large number of equations present (1 for the bosonic field plus $N_F$ for the fermionic fields) make the numerical analysis of realistic systems having  thousands of fermions, unfeasible in multidimensional configurations.

Boson-Fermion mixtures as modeled by Eqs. (\ref{NLSx}) are also a particular case of our general model for multicomponent NLS systems given by Eqs. (\ref{nlse-multi}).

Finally we would like to mention on yet another application of our model equations in the limit of a large number of components (formally $M \rightarrow \infty$), which is that of white light propagation in nonlinear media \cite{Segev}.

In this paper we study Eqs. (\ref{nlse-multi}) to obtain  sufficient conditions for blow-up and for the existence of global in time solutions in two and three spatial dimensions. Our first step is to develop the theory based on the moment method (Sec. \ref{MII}). To do so  we first define the relevant quantities for our analysis and find their evolution equations (Sec. \ref{II}). Next, the equations related to the global motion of the solution are studied in Sec. \ref{III} while those related to the width  are studied in Sec. \ref{IV}. The results on sufficient conditions for blow-up are then presented in Sec. \ref{suffi}. Finally, in Sec.  \ref{V} we consider the problem of finding sufficient conditions for the existence of global solutions. Our conclusions are then summarized in Sec. \ref{VI}.

\section{Moment equations}
\label{MII}

\subsection{Definition and evolution of the moments}
\label{II}

Our goal is to obtain a suficient condition for collapse of initial data $\psi_k(\boldsymbol{x},t=0,)$ satisfying equations (\ref{nlse-multi}).
  In what follows we will take the potentials to be of the form
\begin{equation}
\label{potential}
 V_k(\boldsymbol{x}) = \frac{1}{2} \omega_{k}^{2} \, \boldsymbol{x}^{2},
 \end{equation}
 which includes the explicit forms of potentials found in realistic applications of NLS to the field of matter waves.
 
The first part of our analysis will be based on an extended version of  the method of moments  \cite{PG1,PG2} for our vector system in order to obtain a virial type identity. The same method in a simpler form (i.e. using a smaller number of moments) has been used by different authors to deal with simpler situations (e.g.  \cite{Berge2,Ghosh,Pomponio,Fanelli}).
 To apply the method we first write $\psi_k$ in the modulus-phase representation 
 \begin{equation}
 \psi_k = \sqrt{n_k} e^{i\varphi_k},
 \end{equation}
  and define the following functionals
\begin{subequations}
\begin{eqnarray}
  N_{k} & = & \myint{  n_{k}} ,
\\
  \mathbf{R}_{k}  & = & \myint{n_{k} \boldsymbol{x}},
\\
  \mathbf{P}_{k} & = & \myint{ n_{k} \nabla\varphi_{k}} =
  \frac{ 1 }{ 2i } 
  \myint{ \left( 
    \bar{\psi}_{k} \nabla\psi_{k} - \psi_{k} \nabla \bar{\psi}_{k} 
  \right)},
\\
  J_{k} & = & \myint{ n_{k} \left( \boldsymbol{x}, \nabla\varphi_{k} \right)} =
  \frac{ 1 }{ 2i } 
  \myint{\left( \boldsymbol{x}, \; 
    \bar{\psi}_{k} \nabla\psi_{k} - \psi_{k} \nabla \bar{\psi}_{k} 
  \right)},
\\
  I_{k} & = & \myint{  n_{k} \, \boldsymbol{x}^{2}},
\\
  N_{kl} & = & \myint{ n_{k} n_{l}},
\\
  K_{k} & = & 
  \myint{ 
    \left( \nabla\psi_{k},  \nabla\bar\psi_{k} \right)},
\\ \label{ham}
  2\mathcal{H} & = & 
    \sum_{k} \lambda_k K_{k} 
  + \sum_{k} \omega^{2}_{k} I_{k} 
  + \sum_{kl} g_{kl} N_{kl},
\end{eqnarray}
\end{subequations}
with $k,\,l=1,...,M$. Hereafter integration is taken over the whole space $\mathbb{R}^{D}$. We will assume that all these quantities are finite for our initial data and are smooth as functions of time.

After direct differentiating the previous equations with respect to time and some rather long but straightforward calculations we get a series of ordinary differential equations for the moments involving different integral quantities
\begin{subequations}
\label{diffmo}
\begin{eqnarray}\label{norma}
\frac{dN_k}{dt} & = & 0, \\ \label{nj}
\frac{d\mathbf{R}_k}{dt}   & = & 
\lambda_k \mathbf{P}_{k},
\\ \label{pj}
\frac{d \mathbf{P}_{k}}{dt}   & = & 
    - \omega_{k}^{2} \mathbf{R}_{k} 
    + \frac{1}{2} \sum_{l} g_{kl} 
      \myint{\left( n_{l} \nabla n_{k} - n_{k}\nabla n_{l} \right)}, \\
\label{dj}
  \frac{d J_{k}}{dt} & = & 
  - \omega_{k}^{2} I_{k}
  + \lambda_k K_{k} 
  - \sum_{l} g_{kl} \myint{ n_{k} \left( \boldsymbol{x}, \nabla n_{l} \right)},
\\
\label{di}
  \frac{dI_{k}}{dt} & = & 
   2\lambda_k J_{k},
\\
  \frac{dK_{k}}{dt} & = & 
  - 2 \omega_{k}^{2} J_{k}
  - 2 \sum_{l} g_{kl} 
      \myint{ n_{k} \left( \nabla\varphi_{k}, \nabla n_{l} \right)}, \\
  \frac{dN_{kl}}{dt} & = & 
  \myint{ \left[
    \lambda_k n_{k} \left( \nabla\varphi_{k}, \nabla n_{l} \right) +
    \lambda_l n_{l} \left( \nabla\varphi_{l}, \nabla n_{k} \right) 
  \right]},
\\ \label{hamiltonian}
  \frac{d\mathcal{H}}{dt} & = & 0.
\end{eqnarray}
\end{subequations}
Eqs. (\ref{norma}) and (\ref{hamiltonian}) express respectively the conservation of the $L^2$-norm and the
Hamiltonian which is natural for NLS-type equations. 

\subsection{Equations for the centroid of the solution}
\label{III}

Eqs. (\ref{nj}) and (\ref{pj}) can be combined to get
\begin{equation}\label{cm}
\frac{1}{\lambda_k} \frac{d^2\mathbf{R}_k}{dt^2} + \omega_k^2 \mathbf{R}_k =  \mathbf{F}_k,
      \end{equation}
where 
\begin{equation}
\mathbf{F}_k =  \frac{1}{2} \sum_{l} g_{kl} 
      \myint{ \left( n_{l} \nabla n_{k} - n_{k}\nabla n_{l} \right)}.
      \end{equation}
     In the case of symmetric interaction matrices and homogeneous systems such as those studied up to now in the literature (i.e. $\omega_k =\omega$, $\lambda_k = \lambda$ for all $k=1,...,M$) these equations are a simple extension of the Ehrenfest theorem for multicomponent systems since one can define $\mathcal{R} = \sum_{k=1}^M \mathbf{R}_k$ which satisfies the average equation $\ddot{\mathcal{R}} + \omega^2 \mathcal{R} = 0$. This equation provides a simple extension of the result for one component \cite{PG2} and manifests the fact that the dynamics of the averaged center of mass behaves independently of internal forces. 
     
In the case of initial data with small overlapping Eqs. (\ref{cm}) can be used as the basis for an approximate theory for the evolution of the centroids of the individual wavepackets following the ideas presented in Ref. \cite{SolMol}. Since in this paper we focus our attention on collapse phenomena we will not perform such analysis here.

\subsection{Equations for the width of the solution}
\label{IV}

From Eqs. (\ref{di}) and (\ref{dj}) we can obtain an equation for the evolution of the width of the individual solutions $\psi_k$ which is parameter used to detect collapse in the framework of moment methods
\begin{equation}
\frac{1}{\lambda_k} \frac{d^2I_{k}}{dt^2} +  2 \omega_{k}^{2} I_{k} = 2 \lambda_k K_{k} 
  - 2 \sum_{l} g_{kl} \myint{n_{k} \left( \boldsymbol{x}, \nabla n_{l} \right)}.
\end{equation}
Summing for all values of $k$ and using the definitions of the momenta we find:
\begin{equation}
\label{rel}
  \sum\limits_{k} \left(
    \frac{1}{\lambda_k} \ddot{I}_{k} + 
    4 \omega^{2}_{k} I_{k} 
  \right) =
  4\mathcal{H} + 
  (D-2) \sum\limits_{kl} g_{kl} N_{kl},
\end{equation}
This equation is the extension of the well known virial relations for the scalar cubic nonlinear Schr\"odinger equation \cite{Sulem,self-focusing} for our multicomponent system ruled by Eqs. (\ref{nlse-multi}).

\section{Sufficient conditions for blow-up}
\label{suffi}

In this section we will use the results of the method of moments to provide sufficient conditions for blow-up of solutions of  equations (\ref{nlse-multi}) with   external potentials the form (\ref{potential}) in different scenarios.

Before moving further we would like to remark that the Hamiltonian $\mathcal{H}$ is a sum of three terms $\mathcal{H}=\mathcal{H}^{(1)}+\mathcal{H}^{(2)}+\mathcal{H}^{(3)}$ such that two of them $\mathcal{H}^{(1)}=\sum\limits_{k}\lambda_k K_k/2$ and $\mathcal{H}^{(2)}=\sum\limits_{k}\omega_k^2 I_k/2$ are always positive. The sign of the third one $\mathcal{H}^{(3)}=\sum\limits_{k}g_{kl}N_{kl}/2$ depends on the properties of the matrix $\{g_{ik}\}$. Let us note that if the matrix $\{g_{kl}\}$ is such that $\mathcal{H}^{(3)}>0$ for any $N_{kl}$ then collapse is not possible. Because of this we will be interested in the class of  matrices $\{g_{kl}\}$ which permits the negativity of the biquadratic form $\mathcal{H}^{(3)}<0$.

The question of the full description of the class of such matrices is closely related to the question of sign definition of the bilinear form $g_{ik}\xi_{i}\xi_{k}$ but not equivalent to it, since in our case we have a strong cone restriction $\xi\geq 0$. We will not study this question in all its generality, but present one sufficient and easily verified characterization of a wide class of matrices for which $\mathcal{H}^{(3)}$ can become negative.

\textbf{Definition} It is said that the matrix $\{a_{ik}\}$ belongs to the class $\mathfrak{N}$ if there is a pair of indices $i_0,\,k_0$ such that
\begin{equation}
  \sqrt{a_{i_0i_0}}\sqrt{a_{k_0k_0}}+a_{{i_0}{k_0}} <0.
\end{equation}

One can easily check that if a matrix $\{g_{kl}\}\in \mathfrak{N}$, then there is a vector $\vec{\xi}$ satisfying the restriction $\xi_i\geq 0$ such that $\sum\limits_{kl}g_{kl}\xi_k\xi_l<0$. For example one can choose 
$\xi_{i_0} = 1/\sqrt{g_{i_0i_0}} $, $\xi_{k_0} = 1/\sqrt{g_{k_0k_0}} $ and $\xi_i=0$ otherwise.

In what follows we will assume that the matrix $g_{lk}\in \mathfrak{N}$ (for example, considering the system (\ref{NLSx}) this means that $g_{BF}<0$) 
and obtain sufficient conditions for blow-up in that situations. In what follows we will treat the cases $D=2,3$ separately.

\subsection{Case $D=2$}

Let us first study the case in which $D=2$.  Eq. (\ref{rel}) becomes
\begin{equation}
\label{diff_eq}
  \sum\limits_{k} 
    \frac{1}{\lambda_k} \ddot{I}_{k}  =  -   \sum\limits_{k}  \omega_k^2 I_k + 4\mathcal{H}.
\end{equation}
Let us define 
\begin{equation}
\mathcal{I}=  \sum\limits_{k} 
 I_{k}/\lambda_k,
\end{equation} 
and
\begin{equation}
\Omega^2 = \min\limits_{k}\left(\lambda_k\omega^2_k\right).
\end{equation}
Since $\lambda_k,\,I_k >0$ by definition, from (\ref{diff_eq}) we get the simple evolution equation
\begin{equation}
  \ddot{\mathcal{I}} +\Omega^2\mathcal{I}=4\mathcal{H}-\delta^2,
  \label{diff_eq_main}
\end{equation}
where
\begin{equation}
 \delta^2 = \sum\limits_{k}\frac{1}{\lambda_k}\left(\lambda_k\omega_k^2-\Omega^2\right)\mathcal{I}_k>0.
\end{equation}
Solving (\ref{diff_eq_main}) we get
\begin{multline}\label{expl}
\mathcal{I}(t)= \left[\mathcal{I}(0)-\frac{4\mathcal{H}}{\Omega^2}\right]\cos (\Omega t)+\frac{\dot{\mathcal{I}(0)}}{\Omega}\sin(\Omega t)+\frac{4\mathcal{H}}{\Omega^2} \\ -\frac{1}{\Omega}\int\limits_{0}^{t}\sin(\Omega(t-\tau))\delta^2(\tau)\,d\tau.
\end{multline}
From Eq. (\ref{expl}) it follows that for all $ t<\pi/\Omega$
\begin{equation}
  \mathcal{I}(t)< \mathcal{I}(0)\cos (\Omega t)+ \frac{\dot{\mathcal{I}(0)}}{\Omega}\sin(\Omega t)+\frac{4\mathcal{H}}{\Omega^2}\left[1-\cos(\Omega t)\right].
  \label{blow-up_ineq}
\end{equation}
This inequality is of the type used in the works \cite{berge97,berge2000,berge2000-2} to study the stability and collapse phenomena in the one-component NLS equations. From this inequality one can conclude that if there are initial conditions such that for some time $T^*\leq \pi/\Omega$ the quantity $I(T^*)$  becomes negative, then there will be blow-up. The complete study of the conditions which can ensure the negativity of the right-hand side of Eq. (\ref{blow-up_ineq}) is a rather difficult problem. Here we present two (in our opinion relevant) examples of the situations when the analysis can be carried out.

\textit{Example 1.}
Let us choose initial conditions such that 
\begin{equation}
\dot{\mathcal{I}}(0)=0,
\label{ex1_1}
\end{equation}
which can be easily done by taking the initial data to be real. Then it follows from Eq. (\ref{blow-up_ineq}) that there exists a $T^*\leq\pi/\Omega$ such that if 
\begin{equation}
\mathcal{H} \leq \frac{1}{8}\Omega^2\mathcal{I}(0),
\label{ex1_2}
\end{equation}
then the solution will blow-up on the time interval $(0,T^*)$.

\textit{Example 2.}
If the initial data are such that
\begin{equation}
 \dot{\mathcal{I}}(0)<0,
 \label{ex2_1}
\end{equation}
what always can be achieved by the appropriate choice of phases $\varphi_k$ of the initial distribution, then if
\begin{equation}
 \mathcal{H} < \frac{1}{4}\Omega|\dot{\mathcal{I}}(0)|,
 \label{ex2_2}
\end{equation}
then the solution blows up for times smaller or equal to $T^*=\pi/(2\Omega)$.

In what follows we will discuss  how to find  initial conditions for our equation (\ref{nlse-multi}) with the given potentials (\ref{potential}) that can fulfill inequalities (\ref{ex1_2}) and (\ref{ex2_2}). 
 Our goal is to show that the class of initial conditions satisfying (\ref{ex1_2}) or (\ref{ex2_2}) is not empty. The proof of this fact is done in two steps.

Step 1. Since $\{g_{kl}\}\in \mathfrak{N}$, then there exists a vector $\xi$ ($\xi_i\geq 0$) such that $g_{kl}\xi_k\xi_l<0$. Choosing $\psi_k(t=0)=\sqrt{\xi_k}\phi(x)$ where $\phi(x)$ is an arbitrary square integrable function we obtain that
\begin{equation}
  \mathcal{H}^{3}[\psi_{k}]<0.
\end{equation}
It is worth noting that such a choice of initial conditions is \textit{not a one-component reduction} in Eqs. (\ref{nlse-multi}) because of the presence of different coefficients $\lambda_k$, $\omega_k$ and $g_{kl}$  (the possibility of such a reduction would imply very strong conditions on the coefficients $\lambda_k$, $\omega_k$, $g_{kl}$ which we do not impose).

Step 2. Since $\mathcal{H}^{3}[\psi_{k}]$ can be done negative it is easy to satisfy the condition (\ref{ex1_2}). To this end we will use a rescaling argument. Let us define
\begin{equation}
 \tilde{ \psi_k}(\boldsymbol{x})=\alpha\psi_k(\boldsymbol{x}), \quad 
  \tilde{\mathcal{H}}= \mathcal{H}[\tilde{\psi_k}], \quad \tilde{\mathcal{I}}=\mathcal{I}[\tilde{\psi_k}].\end{equation}
Then
\begin{equation}
\tilde{\mathcal{H}} - \frac{1}{8}\Omega^2\tilde{\mathcal{I}}(0)=
\alpha^2\left(\mathcal{H}^{(1)}+\mathcal{H}^{(2)}-\frac{1}{8}\Omega^2\mathcal{I}(0)-\alpha^2|\mathcal{H}^{(3)}|\right).
\label{comp_cond_resc}
\end{equation}
From the last relation it follows directly that one can always choose $\alpha$ sufficiently large to make (\ref{comp_cond_resc}) negative and thus to satisfy (\ref{ex1_2}).

The case of condition (\ref{ex2_2}) is verified in a similar way.


\textbf{Remark} Here we would like to make a remark concerning one very relevant from the physical point of view situation. Let us consider the system of equations (\ref{nlse-multi}) in the particular case when $M=2$
\begin{subequations}
\label{nlse-multi2}
\begin{equation}
  i\partial_{t} \psi_{1} = 
  - \frac{1}{2}\lambda_1 \Delta \psi_{1} + V_1 \psi_1 +  \left( \sum\limits_{l=1}^{2} g_{1l} \left|\psi_{l}\right|^{2} \right) \psi_1,
\end{equation}
\begin{equation}
  i\partial_{t} \psi_{2} = 
  - \frac{1}{2}\lambda_2 \Delta \psi_{2} + V_2 \psi_2 +  \left( \sum\limits_{l=1}^{2} g_{2l} \left|\psi_{l}\right|^{2} \right) \psi_2.
\end{equation}
\end{subequations}
From the previous considerations it follows  that if the matrix $\{g_{kl}\}_{k,l=1}^{2}$ belongs to the class $\mathfrak{N}$ then the solution can blow-up in finite time. For the particular case of $M=2$ it follows from the definition of $\mathfrak{N}$ that for $\{g_{kl}\}_{k,l=1}^{2}\in \mathfrak{N}$ it is sufficient that
\begin{equation}
  g_{12}<-\sqrt{g_{11}g_{22}}.
  \label{M2Ndef}
\end{equation}
This means that even in the case of repulsive intracomponent action terms ($g_{11},\, g_{22}>0$)  blow-up phenomena can develop in finite time due to the presence of an attractive intercomponent interaction satisfying ${g_{11}g_{22}}<g^2_{12}$. Moreover if $g_{11},g_{22}>0$ and Eq. (\ref{M2Ndef}) does not hold, then there are no blowing-up solutions of Eq. (\ref{nlse-multi2}).

\subsection{Case $D=3$.}

The same type of estimates can be applied to the situation when the matrix $\{g_{kl}\}$ is negative definite and $D=3$ since then being $N_{kl} >0$ we can write
 \begin{equation}
\label{rel2}
  \ddot{\mathcal I}  =  -   \sum\limits_{k}  \omega_k^2 I_k +   \sum\limits_{kl} g_{kl} N_{kl} +  4\mathcal{H},
\end{equation}
and thus one can obtain a similar to the two-dimensional case  sufficient condition for the existence of blow-up. The same comment done in the previous subsection applies here, i.e. one may think of situations where a single negative interaction coefficient is able to drive the full system to blow-up.


\section{Global in time solutions}
\label{V}

In this section, assuming that the matrix $\{g_{kl}\}$ has at least one negative element, we will obtain conditions that should be imposed on the initial data in dimensions, $D=2,3$ to guarantee that the solution is uniformly bounded for all times, i.e. we will show that if the initial data are sufficiently small then there 
is no blow-up in the framework of Eqs. (\ref{nlse-multi}). The case when all the elements of the matrix $\{g_{kl}\}$ are positive is trivial since the global solvability  follows directly from the structure of the Hamiltonian. Moreover we will suppose that $\mathcal{H}>0$, since the case of $\mathcal{H}<0$ implies that the initial conditions should be sufficiently ''large'' \cite{Cazenave} (this is the consequence of the rescaling $\tilde{\psi_k}=k\psi_k$ and the structure of the Hamiltonian).

\subsection{D=3}
Let us first write the conservation of energy in the general form:
\begin{equation}
    2\mathcal{H} = \sum\limits_{k}\lambda_k \myint{ |\nabla \psi_k|^2} + \sum\limits_{kl}g_{kl}\myint{ |\psi_k|^2|\psi_l|^2}+\sum\limits_{k}\myint{ V_k|u_k|^2}.
\end{equation}
To obtain some useful estimations in the case of $D=3$ we will now derive a multidimensional Gagliardo-Nirenberg inequality, namely
\begin{eqnarray}
  \sum\limits_{kl}\Vert \psi_k\psi_l  \Vert^2_{L^2}
  \leq
  MC^{(3)}_{\scriptscriptstyle GN}\mathcal{N}^{1/2}\left( \sum\limits_{k}\Vert \nabla \psi_k \Vert^2_2\right)^{3/2},
  \label{GN}
\end{eqnarray}
here by $\Vert\cdot\Vert_{L^p}$ we define the $L^p$ norm : $\displaystyle{\Vert v\Vert_{L^p} =\left( \myint{|v|^p}\right)^{1/p}}$,  $C^{(3)}_{\scriptscriptstyle GN}$ stands for the best constant in the three dimensional scalar Gagliardo-Nirenberg inequality \cite{DelPino}, $M$ is the number of components in our system and $\mathcal{N}=\sum\limits_{k}\Vert \psi_k \Vert_{2}^2$. Hereafter we present a simple algebraic proof. Let us consider the Cauchy-Bunyakovsky inequality :
\begin{equation}
    \sum\limits_{kl}\int\limits \psi_k^2\,\psi_l^2\,d^{\scriptscriptstyle 3}\!\vec{x} \leq M \sum\limits_{k} \myint{\psi_k^4}.
    \label{Helder}
\end{equation}
Now, let us apply the standard one component  Gagliardo-Nirenberg inequality for spatial dimension $D=3$~\cite{Cazenave}
\begin{equation}
    \norm{v}_{L^4}^4\leq C^{(3)}_{\scriptscriptstyle GN}\norm{\nabla v}_{L^2}^3\norm{v}_{L^2},
\end{equation}
where. Then from (\ref{Helder}) we obtain that
\begin{equation}
    \sum\limits_{kl}\int \psi_k^2\,\psi_l^2\,d^{\scriptscriptstyle 3}\!\vec{x} \leq M C^{(3)}_{\scriptscriptstyle GN}
    \sum\limits_{k}\norm{\nabla \psi_k}_{L^4}^3\norm{\psi_k}_{L^2},
\end{equation}
Thus applying H\"older inequality one more time we obtain the needed inequality
\begin{equation}
      \sum\limits_{kl}\int \psi_k^2\,\psi_l^2\,d^{\scriptscriptstyle 3}\!\vec{x} \leq M C^{(3)}_{\scriptscriptstyle GN} \mathcal{N}^{1/2} \left(\sum\limits_{k}\norm{\nabla \psi_k}_{L^2}^{2}\right)^{3/2},
      \end{equation}
which corresponds to the result presented in inequality (\ref{GN}). 
From this inequality and the conservation of energy we can easily obtain that:
\begin{eqnarray}
   2\mathcal{H} \geq \lambda_{min}\sum\limits_{k}\Vert
   \nabla \psi_k \Vert_{L^2}^2 - g_{*}MC^{(3)}_{\scriptscriptstyle GN} \sqrt{\mathcal{N}}\left(\sum\limits_{k}\Vert \nabla \psi_k \Vert^2_{L^2}\right)^{3/2},
\end{eqnarray}
where by $\lambda_{min}$ we denote the minimal value of all $\lambda_k$ and by $g_*$ we denote the modulus of the minimal element of the matrix $\{g_{kl}\}$.
Let us define 
\begin{equation}
 U^2= \sum\limits_{k}\Vert
   \nabla \psi_k \Vert_2^2.
\end{equation}
Then
\begin{equation}
    2\mathcal{H} \geq \lambda_{min} U^2- \gamma\sqrt{\mathcal{N}}U^3,
\end{equation}
where $\gamma= g_*MC^{(3)}_{\scriptscriptstyle GN} $. Define
\begin{equation}
   F(U)=\gamma\sqrt{\mathcal{N}}U^3-\lambda_{min} U^2+2\mathcal{H}\geq 0.
\end{equation}
The function $F(U)$ is a cubic polynomial of $U$, such that $F(0)= 2\mathcal{H}\geq 0$ with two critical points $U_0=0$ and $U_1 = (2/3)({\lambda_{min}}/{\gamma\sqrt{\mathcal{N}}})$. Since for any sufficiently smooth initial the solution is locally smooth in time, then if we choose the initial conditions satisfying the following conditions

\begin{itemize}
\item $F(0)>0,$
\item $F(U_1)\leq 0,$
\item $U(t=0)$ before the first positive root of $F(U)=0$.
\end{itemize}

then the corresponding solution is globally bounded in time.  

The first condition is satisfied automatically since $\mathcal{H}>0$. As 
\begin{equation}
    F(U_1)= - \frac{4}{27}\frac{\lambda_{min}^3}{\gamma^2\mathcal{N}} +2\mathcal{H},
\end{equation}
then the second condition is equivalent to the 
\begin{equation}
  \mathcal{N}  |\mathcal{H}|\leq \frac{2}{27}\frac{\lambda_{min}^3}{\gamma^2}.
  \label{restr1}
\end{equation}
Noting that $F(\sqrt{2\mathcal{H}/\lambda_{min}})\geq0$ and since $\sqrt{2\mathcal{H}/\lambda_{min}} \leq U_1$ one can conclude that to satisfy the third condition it is sufficient to choose
\begin{equation}
  U(t=0)\leq \sqrt{\frac{2\mathcal{H}}{\lambda_{min}}}.
  \label{restr2}
\end{equation}

The conditions (\ref{restr1}) and (\ref{restr2}) are compatible for any choice of parameters $\lambda_k,\, \omega_k$ and $g_{ik}$ which is easily  verified using the rescaling $\tilde{u}(y)=\alpha u(\beta x)$.

\subsection{$D=2$}

Now let us analyze the case $D=2$. The same ideas used to derive  (\ref{GN}) allow to obtain that if $D=2$ then:
\begin{equation}
    \sum\limits_{kl}\int \psi_k^2\,\psi_l^2\,d^{\scriptscriptstyle 2}\!\vec{x} \leq 
    M\mathcal{N}C^{(2)}_{\scriptscriptstyle GN}\sum\limits_{k}\norm{\nabla \psi_k}_{L^2}^2,
\end{equation}
where $C^{(2)}_{\scriptscriptstyle GN}$ is the optimal constant for Gagliardo-Nirenberg inequality in dimension two.
So from (\ref{rel2}) we obtain that:
\begin{eqnarray}
2\mathcal{H} \geq \left(\lambda_{min}-g_*M \mathcal{N}C^{(2)}_{\scriptscriptstyle GN}\right)\sum\limits_{k}\norm{\nabla \psi_k}^2_{L^2}.
\end{eqnarray}
It follows from the last inequality and the constancy of the Hamiltonian, that if we chose initial conditions such that
\begin{equation}
\mathcal{N}= \sum\limits_{k}\norm{\psi_k}^2_{L^2}\leq\dfrac{\lambda_{min}}{g_*MC^{(2)}_{\scriptscriptstyle GN}},
\end{equation}
then the solution is also globally bounded in time.

Thus we have proved the following theorem
\begin{theorem}
If the initial conditions for system of equations (\ref{nlse-multi}) with the potentials of the form (\ref{potential}) and with the matrix $\{g_{kl}\}$ having at least one negative element are such that $\mathcal{H}<0$ and
\begin{itemize}
\item
D=2 : 
\begin{equation}
\mathcal{N}= \sum\limits_{k}\norm{\psi_k}^2_{L^2}\leq\dfrac{\lambda_{min}}{g_*MC^{(2)}_{\scriptscriptstyle GN}},
\end{equation}

\item 
D=3:
\begin{equation}
  \mathcal{N}  |\mathcal{H}|\leq \frac{2}{27}\frac{\lambda_{min}^3}{\gamma^2}.
  \end{equation}
\begin{equation}
  U(t=0)\leq \sqrt{\frac{2\mathcal{H}}{\lambda_{min}}}.
\end{equation}

\end{itemize}
then the corresponding solution is globally bounded in time.
\end{theorem}
\section{Conclusions}
\label{VI}

In this paper we have studied system of coupled nonlinear Schr\"o\-dinger equations with potentials which include as particular cases models of heteronuclear multicomponent Bose-Einstein condensates and mathematical models for Boson-Fermion mixtures. For these models we have stabilished different types of results on blow-up.

First we have obtained sufficient conditions for blow-up using an extension of the method of moments valid for two or three spatial dimensions. 
In the most interesting situations where the coefficients of the interaction matrix are not all of the same sign our criterion allows to predict that there will be situations when the system can be driven to blowup either by a single unstable component or because of negative cross-interaction coefficients.

As a second class of results we have found precise sufficient conditions for the initial data ensuring the existence of global solutions which makes explicit the intuitive idea that small enough data should be non-collapsing.

\textbf{Acknowledgements}

We want to acknowledge Vladimir Konotop (U. of Lisbon) for many fruitful discussions on this topic. This work has been supported by grants: FIS2006-04190
(Ministerio de Educaci\'on y Ciencia, Spain), PCI08-0093 (Consejer\'{\i}a de Educaci\'on y Ciencia
de la Junta de Comunidades de Castilla-La Mancha, Spain). V. Prytula is supported by grant : AP2005--4528 (Ministerio de Educaci\'on y Ciencia, Spain)


\begin{thebibliography}{99}

\bibitem{Berge}{L. Berg\'e, Wave collapse in Physics: Principles and applications to light and plasma waves,
Phys. Rep. \textbf{303},  259-370 (1998).}

\bibitem{Sulem}{C. Sulem and P. Sulem, ``The nonlinear Schr\"{o}dinger equation: Self-focusing and wave collapse", Springer, Berlin (2000).}

\bibitem{self-focusing} V. I. Bespalov, and V. I. Talanov,
ZhETF Pis'ma {\bf 11}, 307 (1966).

\bibitem{Kelley}{P.L. Kelley, Self focusing of optical beams, Phys. Rev. Lett. \textbf{15}, 1005 (1965).}

\bibitem{Hulet}{C. C. Bradley, C. A. Sackett, J. J.Tollett, and R. G. Hulet, Evidence of Bose-Einstein Condensation in an Atomic Gas with Attractive Interactions, Phys. Rev. Lett. {\bf 75},  1687  (1995).}

\bibitem{Hulet2}{C. C. Bradley, C. A. Sackett, and R. G. Hulet, Bose-Einstein Condensation of Lithium: Observation of Limited Condensate Number, Phys. Rev. Lett. \textbf{78}, 985 (1997).}

\bibitem{F3}{E. A. Donley, N. R. Claussen, S. L. Cornish, J. L. Roberts, E. A. Cornell, and C. E. Wieman, Dynamics of collapsing and exploding Bose-Einstein condensates, Nature \textbf{412}, 295 (2001).}

\bibitem{Castella}{F. Castella, $L^2$-solutions to the Schršdinger-Poisson System : Existence, Uniqueness, 
Time Behaviour, and Smoothing Effects, Math. Mod. Meth. Appl. Sci., \textbf{8}, 1051-1083 (1997).}

\bibitem{Pry}  V. Prytula, V. V. Konotop, V. M. P\'erez-Garc\'{\i}a, and V. Vekslerchik,  Collapse in boson-fermion mixtures with all-repulsive interactions, Phys. Rev. A \textbf{76}, 011803 (2007).

\bibitem{Fibich1}{G. Fibich and B. Ilan, Deterministic vectorial effects lead to multiple filamentation, 
Opt. Lett. \textbf{26},  840-842 (2001).}

\bibitem{Fibich2}{G. Fibich and B. Ilan,
Vectorial and random effects in self-focusing and in multiple filamentation, Physica D \textbf{157},  112-146 (2001).} 
 
\bibitem{Fibich3}{G. Fibich and B. Ilan, Multiple filamentation of circularly polarized beams, 
Phys. Rev. Lett. \textbf{89}, 013901 (2002).}

\bibitem{Besse1}{C. Besse and B. Bidegaray, Numerical study of self-focusing solutions to the Schršdinger-Debye system,  Math. Model. Numer. Anal., \textbf{35}, 35-55  (2001).}

\bibitem{Besse2}{C. Besse and C.H. Bruneau, Numerical study of elliptic-hyperbolic Davey-Stewartson system: dromions simulation and blow-up, Math. Mod. and Meth. in Appl. Sciences, \textbf{8}, 1363 (1998).}

\bibitem{McKinstrie}{C. J. McKinstrie, and D. A. Rusell, Nonlinear focusing of Coupled waves, Phys. Rev. Lett. \textbf{61},  2929 (1988).}

\bibitem{Bang}{O. Bang, L. Berg\'e, and J. J. Rasmussen, Fusion, collapse and stationary bound states of incoherently coupled waves in bulk media, Phys. Rev. E, \textbf{59}, 4600 (1998).}

\bibitem{Berge2}{L. Berg\'e, Coalescence and instability of copropagating waves, Phys. Rev. E, \textbf{58}, 6606 (1998).}

\bibitem{Ghosh}{P. K. Gosh, Exact results on the dynamics of a multicomponent Bose-Einstein condensate 
Phys. Rev. A \textbf{65}, 053601 (2002).}

\bibitem{Pomponio}{A. Pomponio, Coupled nonlinear Schr\"odinger systems with potentials,  J. Diff. Equat. \textbf{227}, 258--281 (2006).}

\bibitem{Fanelli} {L. Fanelli and E. Montefusco, On the blow-up threshold for wave weakly coupled nonlinear Schr\"odinger eqations, J. Phys. A: Math. Theor. \textbf{40}, 14139 (2007).}

\bibitem{Modugno}{G. Modugno, M. Modugno, F. Riboli, G. Roati, and M. Inguscio, Two Atomic Species Superfluid, 
Phys. Rev. Lett. \textbf{89}, 190404 (2002).}

\bibitem{Symbiotic}{V. M. P\'erez-Garc\'{\i}a, and J. Belmonte, Symbiotic solitons in heteronuclear multicomponent
Bose-Einstein condensates, Phys. Rev. A \textbf{72}, 033620 (2005).}

\bibitem{col1}{R. Roth, Structure and stability of trapped atomic boson-fermion mixtures, Phys. Rev. A \textbf{66} 013614 (2002).}

\bibitem{col2}{T. Miyakawa, T. Suzuki, and H. Yabu, Induced instability for boson-fermion mixed condensates of alkali-metal atoms due to the attractive boson-fermion interaction, Phys. Rev. A \textbf{64}, 033611 (2001).}

\bibitem{col3}{M. Modugno, F. Ferlaino, F. Riboli, G. Roati, M. Modugno, and M. Inguscio, Mean-field analysis of the stability of a K-Rb Fermi-Bose mixture, Phys. Rev. A \textbf{68}, 011601 (2003).}

\bibitem{karpiuk}{T. Karpiuk, M. Brewczyk, S. Ospelkaus-Schwarzer, K. Bongs, M. Gajda, and K. Rzewski, Soliton Trains in Bose-Fermi Mixtures, Phys. Rev. Lett. \textbf{93}, 100401 (2004).}
 
 \bibitem{salerno}{M. Salerno, Matter wave quantum dots and anti-dots in ultracold atomic Bose-Fermi mixtures, Phys. Rev. A \textbf{72}, 063602 (2005).}
 
 \bibitem{Maruyama}{T. Maruyama, H. Yabu, T. Suzuki, Monopole oscillations and dampings in Boson and Fermion mixture in the time-dependent Gross-Pitaevskii and Vlasov equations, Phys. Rev. A \textbf{75}, 13609 (2005).}
 
 \bibitem{solibf}{J. Belmonte-Beitia, V. M. P\'erez-Garc\'{\i}a, and V. Vekslerchik, Modulational instability, solitons and periodic
waves in a model of quantum degenerate
bosonÐfermion mixtures, Chaos, Solitons and Fractals \textbf{32}, 1268  (2007).}
 
 \bibitem{Segev}{H. Buljan, M. Segev, M. Soljacic, N. K. Efremidis, and D. N. Christodoulides, White-light solitons, Opt. Lett. \textbf{14}, 1239 (2003).}
 
 \bibitem{PG1}{J. J. Garc\'{\i}a-Ripoll, V. M. P\'erez-Garc\'{\i}a, and P. Torres, Extended Parametric Resonances in Nonlinear Schr\"odinger Systems, Phys. Rev. Lett. \textbf{83}, 1715 (1999).}
 
 \bibitem{PG2}{V. M. P\'erez-Garc\'{\i}a, P. Torres, and G. D. Montesinos, The method of moments for Nonlinear Schr\"odinger Equations: Theory and Applications, SIAM J. Appl. Math. \textbf{67}, 990-1015 (2007).}
 
 \bibitem{SolMol}{V. M. P\'erez-Garc\'{\i}a, and V. Vekslerchik, Soliton molecules in trapped vector nonlinear Schr\"odinger systems, Phys. Rev. E \textbf{67}, 061804 (2003).}
 
 \bibitem{Nehari}{T.-C. Lin and J. Wei, Symbiotic bright solitary wave solutions of coupled nonlinear Schr\"odinger equations, Nonlinearity \textbf{19}, 2755-2773 (2006).}

\bibitem{berge97}
{L. Berg\'e, Self-focusing dynamics of nonlinear waves in media with parabolic-type inhomogeneities, Phys. Plasmas, \textbf{4}, 1227 (1997).}

\bibitem{berge2000} {L. Berg\'e, Soliton stability versus collapse, Phys. Rev. E, \textbf{62}, R3071 (2000) }

\bibitem{berge2000-2} {L. Berg\'e, et al., Phys. Rev. A, \textbf{62}, 023601 (2000)} 


\bibitem{Cazenave}{T. Cazenave, An Introduction to Nonlinear Schr\"odinger Equations, Instituto de Matem\'atica, Rio de Janeiro (1996).} 


\bibitem{DelPino}{M, Del Pino and J Dolbeaut, Best constants for Gagliardo-Nirenberg inequalities and applications to nonlinear diffusions,  J. Math. Pures Appl. \textbf{81}, 847-875 (2002).} 



 \end{thebibliography}
\end{document}